# The Adiabatic Piston and the Second Law of Thermodynamics

Bruno Crosignani[*], Paolo Di Porto[**], and Claudio Conti[***]

[*]*Department of Applied Physics, California Institute of Technology, Pasadena, CA 91125*
[**]*Dipartimento di Fisica, Università dell'Aquila, 67010 L'Aquila, Italy and Istituto Nazionale per la Fisica della Materia, Unità di Roma "La Sapienza", 00185 Roma, Italy*
[***]*Istituto Nazionale per la Fisica della Materia, Unità di Roma Tre, 00146 Roma, Italy*

**Abstract.** A detailed analysis of the adiabatic-piston problem reveals peculiar dynamical features that challenge the general belief that isolated systems necessarily reach a static equilibrium state. In particular, the fact that the piston behaves like a *perpetuum mobile*, i.e., it never stops but keeps wandering, undergoing sizable oscillations, around the position corresponding to maximum entropy, has remarkable implications on the entropy variations of the system and on the validity of the second law when dealing with systems of mesoscopic dimensions.

The adiabatic-piston problem is a peculiar example in thermodynamics [1], which has recently been the object of renewed interest [2],[3]. We refer to the model case of an adiabatic cylinder divided into two regions A and B by a movable, frictionless, perfectly insulating piston. Both A and B contain an equal amount of the same perfect gas. We assume the piston to be held up to time t=0 by latches, so that the gases in A and B are initially characterized by well-defined equilibrium states, respectively corresponding to temperatures and volumes $T_A = T_1(0)$, $V_A = SX(0)$ and $T_B = T_2(0)$, $V_B = S[L-X(0)]$, where S and L are the cylinder transverse area and length and X(0) is the initial piston position (see Fig.1).

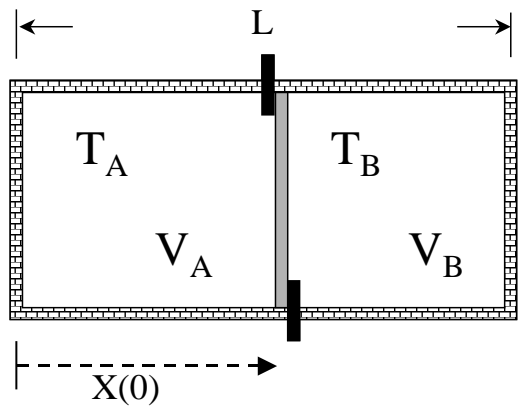

**FIGURE 1.** The adiabatic piston.

Once the latches are released, the piston starts moving and its dynamical evolution is described by means of a suitable kinetic model (see Eqs. (6), (7), (8) of [2]). In particular, the solution of this set of coupled equations allows one to predict the final position reached by the piston for large values of time, as well as the final temperatures and the common value of the final pressure, quantities which are not entirely determinate in the frame of elementary thermodynamics [1]. The model considered in [2] describes situations in which the macroscopic pressure difference in A and B constitutes the main driving force on the piston, so that microscopic pressure fluctuations are consistently neglected.

We wish here to investigate the different regime associated with the initial conditions corresponding to $X(0)=L/2$, $\dot{X}(0)=0$ and to a common value of the initial pressure, and thus of the initial temperature, in A and B. In this situation, the fluctuating microscopic Langevin force, arising from the random nature of the collisions between the piston and the molecules on both sides, cannot be neglected. A preliminary investigation of this regime [4] has revealed the presence of intriguing features. In particular, it has been shown that the piston, initially set in the symmetric position corresponding to equal temperatures on both sides and to maximum entropy of the gaseous system, undergoes a stochastic motion associated with relevant asymptotic fluctuations of the random variable [X(t)-L/2]. As a consequence, the system exhibits negative entropy variations which can be much larger than those associated with standard thermal fluctuations. In the present paper, this surprising result is confirmed and put on a firmer ground by numerically investigating the stochastic motion of the piston after deriving the Langevin force by means of a simple application of the fluctuation-dissipation theorem.

We start by recalling that X(t) obeys the non-linear stochastic equation (see Eq.(7) of [4])

$$\frac{d^2X}{dt^2} = -\frac{(16nRT_0 M_g)^{1/2}}{(\pi LM^2)^{1/2}}(\frac{1}{\sqrt{X}}+\frac{1}{\sqrt{L-X}})\frac{dX}{dt} + \frac{M_g}{M}(\frac{1}{X}-\frac{1}{L-X})(\frac{dX}{dt})^2 + A(t) , \qquad (1)$$

where n is the common mole number in A and B, R is the gas constant, $T_0$ the common value of temperature in A and B when the piston is in the central position X=L/2, $M_g$ the common value of the gas mass in A and B, M the piston mass and A(t) labels the Langevin acceleration. In order to study Eq.(1), we need first to determine the amplitude of A(t).

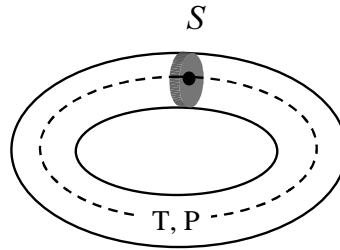

**FIGURE 2.** Geometry of a torus of section S, filled with a gas at temperature T and pressure P.

To this aim, we introduce a configuration slightly modified with respect to that sketched in Fig.1. More precisely, we consider a torus of section S (see Fig.2), filled with a perfect gas at thermal equilibrium at temperature T and pressure P, in which an adiabatic piston of mass M is free to slide without friction under the action of the gas. In this configuration, the equation of motion of the piston is readily obtained in the form

$$M\frac{d^2X}{dt^2} = 2m<[|v_x|-\frac{dX}{dt}]^2>\frac{\rho}{2}S - 2m<[|v_x|+\frac{dX}{dt}]^2>\frac{\rho}{2}S + F(t)$$
$$= -4m<|v_x|>\rho S\frac{dX}{dt} + F(t) , \qquad (2)$$

where m is the molecular mass, $v_x$ the random molecular velocity orthogonal to the piston, $\rho$ the gas number density and F(t)=MA(t) the Langevin force exerted by the gas on the piston. We note that the terms containing $|v_x|$-dX/dt and $|v_x|$+dX/dt represent the pressure forces respectively exerted by the gas on the left and right side of the piston (see Fig.2), and are derived as a simple extension of ordinary kinetic theory to include the effect of finite piston velocity. Equation (2) can be rewritten as

$$\frac{d^2X}{dt^2} = -4(2m/\pi M^2 K_B T)^{1/2} PS\frac{dX}{dt} + A(t) \equiv -\frac{1}{MB}\frac{dX}{dt} + A(t), \qquad (3)$$

where use has been made of the relations $<|v_x|>=(2K_BT/\pi m)^{1/2}$ (an elementary consequence of Maxwell's velocity distribution law) and $\rho=P/K_BT$. Equation (3) is the Langevin equation describing the Brownian motion of the piston inside the torus, and the standard approach allows one to determine the quantity C defined by the relation $<A(t)A(t')>=C\delta(t-t')$. One has, in our one-dimensional case, $C=2K_BT/M^2B$ (see, e.g., [5]), so that

$$<A(t)A(t')> = [8(2mK_BT/\pi)^{1/2}PS/M^2]\delta(t-t'). \qquad (4)$$

We assume Eq.(4) to represent the statistical properties of Langevin's acceleration in our case, described by Eq.(1). In particular, the white-noise behavior is justified by the extremely rapid variations of A(t) compared to those of dX/dt. Therefore, by using Eq.(4) and introducing the characteristic time $t_p = ML(\pi/16nRT_0M_g)^{1/2}$ and the dimensionless variables $\xi=X/L$ and $\tau=t/t_p$, Eq.(1) can be recast in its final form

$$\ddot{\xi} + (\frac{1}{\sqrt{\xi}} + \frac{1}{\sqrt{1-\xi}})\dot{\xi} - \frac{1}{\mu}(\frac{1}{\xi} - \frac{1}{1-\xi})\dot{\xi}^2 = \sigma\alpha(\tau) , \qquad (5)$$

where the dot stands for derivative with respect to $\tau$, $\mu=M/M_g$, $\sigma^2=(\pi/4\sqrt{2})\mu/N$ , N being the common number of gas molecules in the two piston-sections, and $\alpha(\tau)$ represents a unitary-power white noise. The stochastic Eq.(5), describing the random motion of the piston, is the central result of this paper.

Before attempting a numerical analysis of Eq.(5), we recall that we have already obtained, under suitable hypotheses, approximate analytical solutions [4]. More precisely, by assuming $|\xi-1/2|\ll 1$, expanding accordingly the terms appearing on the left-hand side of Eq.(5) and approximating the square of the piston velocity $\dot{X}^2$ with its average thermal value $KT_0/M$, we have been able to approximate Eq.(5) with that describing the Brownian motion of a harmonically bound particle of mass M, a system extensively discussed in the literature [6]. In particular, we have found that, for $\mu<1$, the piston reaches the asymptotic normalized mean-square average position $<(\xi-1/2)^2> \cong \mu/2$ in a time $t_{as}\cong Nt_p$. This implies that the piston does not stop, but keeps wandering around its initial position $\xi=1/2$ undergoing random oscillations whose amplitude is a sizable fraction of the cylinder length.

The results of the numerical analysis of Eq.(5) essentially agree with those worked out in [4], and allows us to extend the investigation of the piston dynamics over a wide range of system parameters. We adopt a second order stochastic leap-frog algorithm as developed in [7] and consider $10^3$ realizations of the piston. As an example, a preliminary analysis of the time evolution of $<(\xi-1/2)^2>$ is reported in Fig.3 for $\sigma=.01$ and $\mu=2$, together with the histogram of the statistical variable $[\xi(\tau)-1/2]^2$ for $\tau=4000$. Its inspection clearly shows how the piston undergoes random fluctuations around the central position $\xi=1/2$, which increase with time up to an asymptotic value corresponding to a sizable fraction of the cylinder length. In terms of dimensional quantities, this specific example corresponds to an asymptotic time $\bar{t} \cong 5000\, t_p \cong 10^4\, L/v_{th}$, where $v_{th}$ is the thermal velocity of the gas molecules.

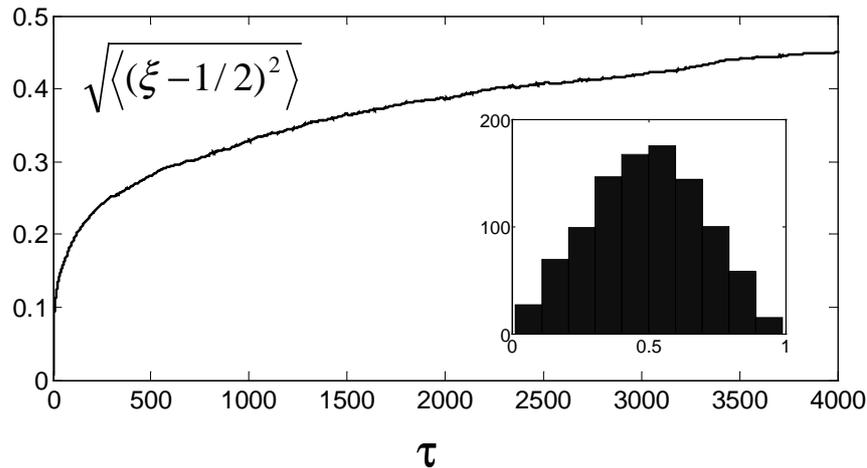

**FIGURE 3.** Time evolution of the normalized root-mean-square deviation of the piston position from its initial value $\xi(0)=1/2$. In the insert, the histogram of the statistical variable $\xi$ for $\tau=4000$.

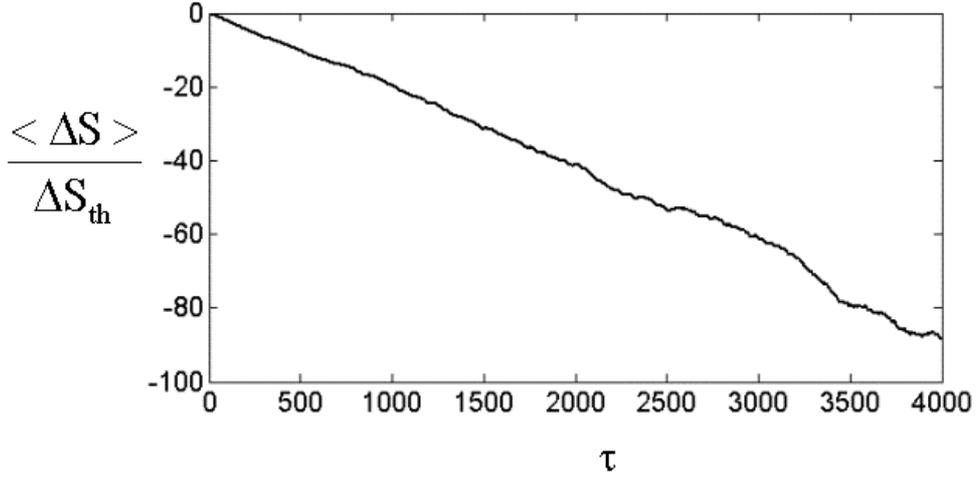

**FIGURE 4.** Time evolution of the normalized entropy variation for σ and μ as in Fig. 3.

Besides, since $\sigma^2 \cong \mu/2N$, our values $\sigma^2=.01$, $\mu=2$ yield $N \cong 10^4$, which for a gas in standard conditions of temperature and pressure implies, for a cubic cylinder, a value of L of the order of a tenth of a micron and an asymptotic time $\bar{t}$ of the order of one microsecond. The entropy variation of the system can be numerically evaluated by taking the average of the general expression ΔS worked out in [4] (see Eq.(5) of [4]), thus obtaining the negative quantity

$$\frac{<\Delta S(t)>}{\Delta S_{th}} = \frac{nc_p}{\Delta S_{th}} < \ln \frac{4LX(t)-4X^2(t)}{L^2} > \cong \sqrt{N} < \ln \frac{4LX(t)-4X^2(t)}{L^2} >, \qquad (6)$$

where use has been made of the relation $\Delta S_{th} \cong (K_B nc_p)^{1/2}$ ($c_p$ representing the gas molar heat at constant pressure) for standard thermal entropy fluctuations [8].

The asymptotic entropy decrease (see Fig. 4) turns out to be larger by two order of magnitude than the corresponding value associated with standard thermal fluctuations, which essentially confirms the violation of the second law in the mesoscopic regime obtained in [4]. We note that this result does not contradict Boltzmann's H-theorem since one of the main hypotheses underlying it, i.e., the molecular-chaos assumption, is not fulfilled by our system [4].

We wish finally to observe that the validity of our model could be fully justified only on the basis of a suitable molecular dynamic simulation involving a considerable number of point particles. Recently, molecular simulations have been carried out, in the frame of the adiabatic piston problem, by modelling the two gases with 500 hard disks separated by a frictionless piston with which they undergo perfect elastic collisions [9]. Actually, the authors do not investigate the behavior of $<X^2(t)>$, but that of $<X(t)>$. More precisely, they consider situations in which $X(0)$ is different from $L/2$ and follow the evolution of the piston along the equal pressure line. In particular, they evaluate the relaxation time over which $<X(t)>$ asymptotically reaches the value $L/2$ (in our case, conversely, $X(0)=L/2$ so that, obviously, $<X(t)>=L/2$ at all times, while $<[X(t)-L/2]^2>$ ranges from the initial value zero to a significant asymptotic value). Their relaxation time turns out to be, as a function of $\mu=M/M_g$, in good qualitative and quantitative agreement with our values of $t_{as}$ [4].

In conclusion, we believe that we have placed into evidence an intriguing aspect of the motion of a frictionless adiabatic piston[1] separating two gases in an adiabatic cylinder. In particular, for mesoscopic time and space scales where the hypotheses underlying our model seem more realistic, the piston never stops but undergoes relevant random fluctuations around its initial position $X(0)=L/2$. As a consequence, this *perpetuum mobile* induces a violation of the second law, in the sense that the entropy of our isolated system decreases a few order of magnitude over standard thermal fluctuations.

## REFERENCES

1. Callen, H. B., *Thermodynamics*, Wiley, New York, 1960, pp. 321-323.
2. Crosignani, B., Di Porto, P., and Segev, M., *Am.J.Phys.* **64**, 610-613 (1996).
3. Gruber, Ch., and Frachebourg, L., *Physica* **272**, 392-428 (1999).
4. Crosignani, B., and Di Porto, P., *Europhys.Lett.***53**, 290-296 (2001).
5. Pathria, R. K., *Statistical Mechanics*, Butterworth-Heinemann, Oxford, 1996, pp. 464-468.
6. Chandrasekhar, S., *Rev.Mod.Phys.***15**, 1-89 (1943).
7. Qiang, J., and Habib, S., *arXiv:physics*/9912055v2 (2000).
8. Landau, L., and Lifshitz, E., *Statistical Physics*, Pergamon Press, Oxford, 1969.
9. Kestemont, E., Van den Broeck, and Malek Mansour, M., *Europhys.Lett*. **49**, 143-149 (2000).

---

[1] The term "adiabatic" requires a clarification. Actually, the random motion undergone by the piston subjects both gases to random impulses. This is in a sense equivalent to permitting heat to flow. Thus, in our case, the term adiabatic only means that the internal degrees of freedom of the piston cannot be excited so that the piston is perfectly reflecting, as well as the side walls of the cylinder